\documentclass[sigconf]{acmart}

\usepackage[boxed]{algorithm2e}
\usepackage{soul}
\usepackage{listings}

\AtBeginDocument{%
  \providecommand\BibTeX{{%
    \normalfont B\kern-0.5em{\scshape i\kern-0.25em b}\kern-0.8em\TeX}}}

\begin{document}

\newcommand{\nb}[2]{
    \fbox{\bfseries\sffamily\scriptsize#1}
    {\sf\small$\blacktriangleright$\textit{#2}$\blacktriangleleft$}
}
\newcommand{\ie}{\textit{i.e.,}~}
\newcommand{\eg}{\textit{e.g.,}~}
\newcommand{\etc}{\textit{etc.}~}
\newcommand{\etal}{\textit{et al.}~}
\newcommand\MICHELE[1]{\textcolor{blue}{\nb{MICHELE}{#1}}}
\newcommand\ALEXEY[1]{\textcolor{red}{\nb{ALEXEY}{#1}}}

\title{\thistool{}: End-to-End Program Repair with LLMs over Retrieval-Augmented Prompts}

\author{Matthew Jin}
\affiliation{%
  \institution{Microsoft}
  \city{Redmond}
  \state{WA}
  \country{USA}
}

\author{Syed Shahriar}
\affiliation{%
  \institution{UCLA}
  \city{Los Angeles}
  \state{CA}
  \country{USA}
}

\author{Michele Tufano}
\affiliation{%
  \institution{Microsoft}
  \city{Redmond}
  \state{WA}
  \country{USA}
}

\author{Xin Shi}
\affiliation{%
  \institution{Microsoft}
  \city{Redmond}
  \state{WA}
  \country{USA}
}

\author{Shuai Lu}
\affiliation{%
  \institution{Microsoft Research}
  \city{Beijing}
  \country{China}
}

\author{Neel Sundaresan}
\affiliation{%
  \institution{Microsoft}
  \city{Redmond}
  \state{WA}
  \country{USA}
}

\author{Alexey Svyatkovskiy}
\affiliation{%
  \institution{Microsoft}
  \city{Redmond}
  \state{WA}
  \country{USA}
}

\newcommand\thistool{InferFix}
\newcommand\inferredbugs{InferredBugs}
\newcommand\TODO[1]{\textcolor{blue}{\textbf{TODO}{#1}}}

\begin{abstract}
Software development life cycle is profoundly influenced by bugs; their introduction, identification, and eventual resolution account for a significant portion of software development cost. This has motivated software engineering researchers and practitioners to propose different approaches for automating the identification and repair of software defects. 

Large language models have been adapted to the program repair task through few-shot demonstration learning and instruction prompting, treating this as an infilling task. However, these models have only focused on learning general bug-fixing patterns for uncategorized bugs mined from public repositories. In this paper, we propose \thistool{}: a transformer-based program repair framework paired with a state-of-the-art static analyzer to fix critical security and performance bugs. \thistool{} combines a Retriever -- transformer encoder model pretrained via contrastive learning objective, which aims at searching for semantically equivalent bugs and corresponding fixes; and a Generator -- a large language model (12 billion parameter Codex Cushman model) finetuned on supervised bug-fix data with prompts augmented via adding bug type annotations and semantically similar fixes retrieved from an external non-parametric memory. 

To train and evaluate our approach, we curated \inferredbugs{}, a novel, metadata-rich dataset of bugs extracted by executing the Infer static analyzer on the change histories of thousands of Java and C\# repositories. Our evaluation demonstrates that \thistool{} outperforms strong LLM baselines, with a top-1 accuracy of 65.6\% for generating fixes in C\# and 76.8\% in Java. We discuss the deployment of \thistool{} alongside Infer at Microsoft which offers an end-to-end solution for detection, classification, and localization of bugs, as well as fixing and validation of candidate patches, integrated in the continuous integration pipeline to automate the software development workflow.

\end{abstract}

\keywords{Program repair, static analyses, prompt augmentation, finetuning}

\maketitle

\section{Introduction}
The software development lifecycle is profoundly affected by bugs. Traditional program analyses techniques can detect and localize bugs through formal reasoning, leaving the task of categorizing the bugs and coding a patch to a developer. The traditional approach of manually generating patches through examination of code changes is a time consuming and error-prone task, which could be automated.

Many of the recently proposed approaches for bug prediction, detection, and repair rely on machine learning algorithms -- infamously \textit{data hungry} --  depending on large amounts of high quality data for effective training. Large language models have been successfully adapted to the program repair tasks through few-shot demonstration learning and instruction prompting, treating this as an infilling task~\cite{ring}. However, while focusing on solving specific research problems they failed to provide a reliable end-to-end program repair solution that could be productized. 

Static analysis tools like Infer can be used to identify critical security and performance issues. This can preempt large parts of the software development cycle, including the process of creating detailed unit tests, which can be extremely time-consuming and difficult for a large, complex project whose code is broken down into many modules or across many files. They can also identify bugs and produce bug reports in a way that is machine-readable and conducive to usage in conjunction with patch generation models.

In this work we focus on three types of bugs reported by Infer: Null Pointer Dereference (NPD), Resource Leak (RL), and Thread Safety Violation (TSV). We focus on these because they pose critical performance, reliability and security issues, and can also be more difficult to fix than other issue types which are also more commonly detected and studied. 

Language models commonly adopt two paradigms for task-specific generalization -- via finetuning or few-shot learning. In the former paradigm, the canonical model training structure is divided into two phases -- pretraining and finetuning. In pretraining stage, a model is trained in a self-supervised way to perform denoising or generic sequence-to-sequence transformations, geared towards improving the performance on a variety of downstream tasks. In the finetuning stage the model is trained on a specialized supervised dataset to perform a concrete task, such as question answering, text summarization, or in our case, program repair. The few-shot learning paradigm allows model specialization for a downstream task via prompt augmentation, composition, or ensembling~\cite{10.1145/3560815}. A variant of prompt augmentation, commonly called $\textit{demonstration learning}$, introduces a few input-output examples for a given task, for instance “The capital of China is Beijing. The capital of Italy is Rome. The capital of South Africa is $\texttt{[X]}$”, allowing to achieve good performance on a downstream task without any gradient updates, which is crucial for very large language models like GPT-3~\cite{gpt3}, T5~\cite{t5}, and PaLM~\cite{palm}. Another variant of prompt augmentation commonly referred to as $\textit{instruction prompting}$ aims to introduce a natural language description of a task, for instance: “write a program to determine whether a graph is bipartite”. It may utilize prompt templates filling in necessary information from an external source (a database, a neural model). In our approach we combine the benefits of both paradigms, by augmenting the prompts and then finetuning our model on the dataset of augmented prompts and predictions to get the best performance.

In this paper, we introduce \thistool -- a program repair framework which combines a transformer encoder model pretrained via contrastive learning serving as a retriever over a database of historic bugs and fixes, and a large language model (12 billion parameter Codex Cushman model, \texttt{code-cushman-001}) instrumented with the facility to leverage retrieved information from the external database. Given the baseline Codex model has been shown to occasionally predict insecure or buggy code~\cite{copilot_bugs}, we prioritized finetuning it on a bug-free supervised dataset of bugs and fixes with contexts enriched via relevant program repair patterns from an external non-parametric memory. The contributions of the paper are as follows: (i) we propose a program repair framework that leverages static analyses for bug detection, localization, and categorization paired with a large language model finetuned for program repair task on a dataset of augmented prompts, (ii) we curate \inferredbugs{}: a metadata-rich dataset of bugs and fixes in Java and C\# programming languages extracted with the Infer static analyzer, (iii) we introduce a dedicated prompt augmentation technique for program repair task, which leverages dense retrieval from an external database of historic bugs and fixes, bug type annotations, and syntactic hierarchies across the entire source code file affected by a bug, (iv) we evaluate our model on the \inferredbugs{} dataset, achieving an impressive 76\% top-1 accuracy of patch generation in Java, and over 65\% in C\#, across null pointer dereference, resource leak, and thread safety violation bug types, and finally (v) we deploy \thistool{} as a GitHub action and as part of the Azure DevOps continuous integration pipeline internally at Microsoft, and document aspects of deployment.

\section{Motivating Example}

To provide the intuition about how our approach works and to describe the concrete details of the bug detection, localization, and repair scenario we begin with a motivating example.

In a typical continuous software development workflow, software engineers make atomic, iterative changes to feature branches periodically merging to the $\texttt{main}$ production branch, which is then continuously and automatically deployed to the end users. Consider a large software project with a modular code base spread across multiple source code files. It can be extremely inefficient, in terms of developer time and effort, to detect, localize, and fix errors manually before they are merged to main. In addition, it requires the creation of an extensive unit test suite to ensure that a feature or change works across all possible versions of the software, and no regressions are introduced. 

\begin{figure}[t]
\includegraphics[width=0.42\textwidth]{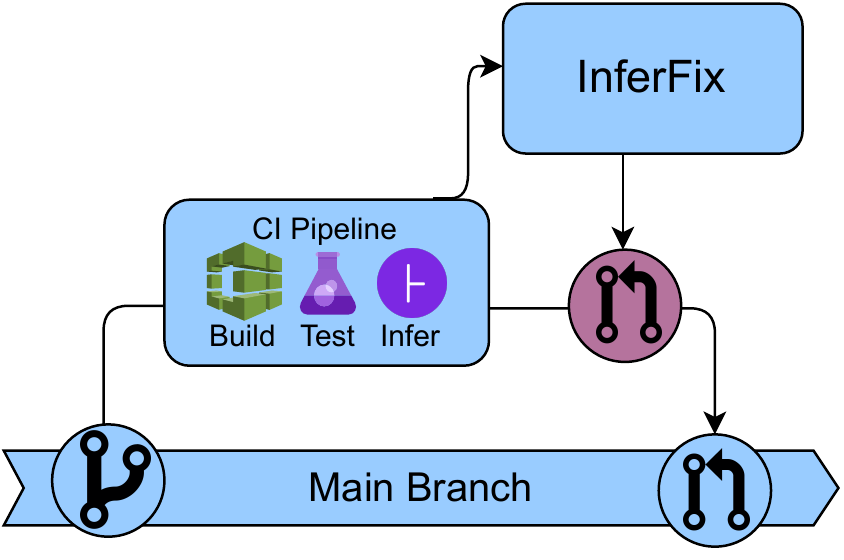}
    \caption{Software development workflow automated with \thistool. A developer creates a pull request to check in code changes implemented in a feature branch, if Infer static analyzer detects a bug, a relevant code context is then prepared and \thistool{} generates a patch which is served as a bug-fixing pull request into feature branch.}
    \vspace{-0.3cm}
    \label{fig:pipeline}
\end{figure}

\autoref{fig:pipeline} illustrates a typical software development workflow at Microsoft Developer Division in presence of \thistool. As a pull request proposing code changes is created, continuous integration pipeline (CI) triggers unit testing, build, and Infer static analysis steps. If bugs are detected, the \thistool patch generation module will be invoked to propose a fix. The proposed bug fix is then validated and subsequently served as a bug-fixing pull request to a feature branch allowing developer to catch bugs before merging the code to the production branch.

Our approach combines a static analyzer to detect, localize, and classify bugs with a powerful LLM (finetuned 12 billion parameter Codex model) to generate fixes.

\begin{figure*}
\includegraphics[width=\textwidth]{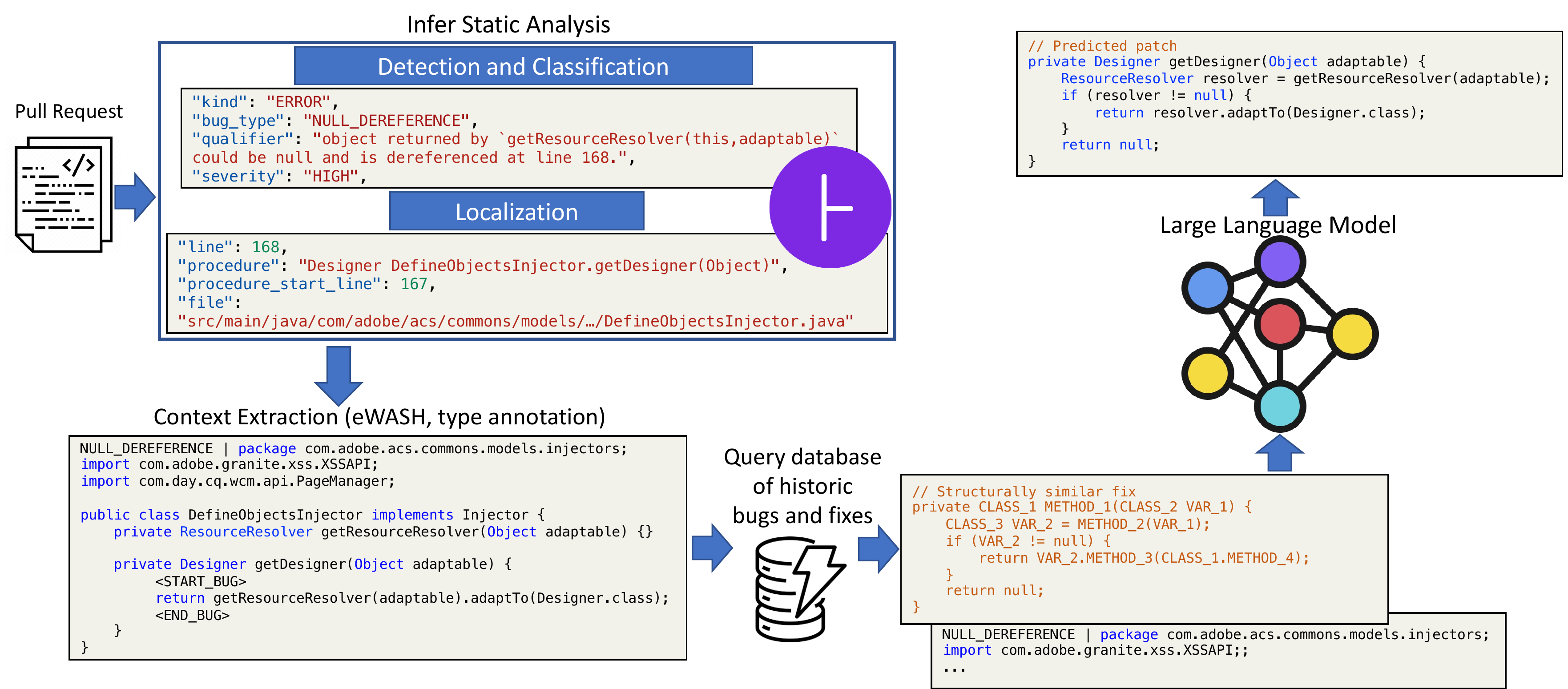}
    \caption{\thistool{} workflow. A buggy commit is detected by the Infer static analyzer, which is utilized to craft a prompt carrying the bug type annotation, location information, relevant syntax hierarchies (eWASH), and similar fixes retrieved from the historic database. A LLM -- finetuned 12B Codex model -- generates a patch.}
    \label{fig:flow}
\end{figure*}

\autoref{fig:flow} provides details about \thistool{} workflow based on a real-world bug example from the acs-aem-common~\cite{acs-aem-common} repository, which is a unified collection of code for content management that optimizes authoring, and delivery of content and digital media written in Java. The Infer static analyzer detects a null pointer dereference error, due to an object in the code returned by $\texttt{getResourceResolver}\\\texttt{(this,adaptable)}$ call, which could be null and is dereferenced at line 168. The context preprocessing module utilizes the information provided by the analyzer to extract the buggy method, and retains surrounding context most relevant to fixing the bug -- import statements, class signature, body of the $\texttt{getResourceResolver}$ method which is invoked at buggy line. The retrieval augmentation engine then searches for semantically similar buggy code snippets in the historic database, prepending similar bug-fixes to the prompt. Finally, the augmented prompt is sent to the finetuned Codex model for inference. The predicted patch is then validated by executing the Infer static analyzer and unit tests as part of the continuous integration pipeline to ensure the error is indeed fixed and no regressions are introduced in the code base.

\section{Dataset}

We collect a supervised dataset of bugs detected with \texttt{Infer} (Infer Static Analyzer), which performs semantic analysis via Separation Logic. 

We executed \texttt{Infer} and \texttt{InferSharp} over the change histories of approximately 6.2k Java and C\# open-source repositories (2.9k Java, 3.3k C\#) hosted on GitHub, analyzing more than 1 million commits. While a few bug datasets are already available, such as Defects4j~\cite{just2014defects4j}, QuixBugs~\cite{lin2017quixbugs}, ManySStuBs4J~\cite{karampatsis2020often}, UnifiedBugDataset~\cite{ferenc2018public} and many others, the dataset we introduce is differentiated by the amount and quality of information provided about each bug by the static analysis. Specifically, each bug in the dataset is associated with several pieces of metadata, including:
\begin{itemize}
    \item \textit{Bug Type}: each detected or fixed bug is marked with a bug type extracted with \texttt{Infer}, such as: null dereference, resource leak, immutable cast, etc. This information could be potentially used by automated program repair techniques to guide the bug-fixing attempts. Alternatively, these instances can be used as labeled data points for bug classification techniques.
    
    \item \textit{Bug Location}: the dataset provides localization info at different levels of granularity: file, class, method, and line. For specific types of bugs, also affected variables/methods are reported.
    
    \item \textit{Change History}: bugs are linked with the change history of the software project. Specifically, the dataset provides information on when a bug was introduced or fixed throughout the development process. Additionally, each analyzed commit is associated with the introduced/fixed or preexisting bugs involving the file touched in the commit.
\end{itemize}

\subsection{Background on Infer Static Analyzer}

Infer is an open-source static analysis tool originating from program analysis research on separation logic. It was first developed by the startup Monoidics Ltd, which was acquired by Facebook in 2013 and open-sourced in 2015. It computes program specifications to detect errors related to memory safety, concurrency, security, and more. It is industrially deployed at companies including Meta, Amazon, and Microsoft. Although in this work we will focus on null dereferences, resource leaks, and thread safety violations detected by Infer, it is able to detect a much wider variety of security and performance issues. For example, via taint tracking it is able to detect dataflow-related issues such as SQL injections. We believe our framework will be capable of mining and generating patches for these bug types as well, but leave the examination of this to future efforts.

At Meta, Infer runs within the internal continuous integration (CI) system of repositories consisting of 10s and 100s millions of lines of code, including those for WhatsApp, Instagram, and Facebook core. Infer runs on diffs and reports issues to developers by writing comments within the code review system. A study conducted at Meta~\cite{meta-study} saw a false positive rate under 20\%, and issues posted saw a fix rate of 70\%. The high issue relevance driven by this diff-time deployment of this system is critical; the same study saw a near-zero fix rate when it was deployed to developers as a list of assigned issues outside of the CI system. This underlines the value of our proposed system being deployed as a bug-detection-and-fix-recommendation code review module.

InferSharp~\cite{infer-sharp} is the compiler frontend developed by Microsoft which translates the Common Intermediate Language (CIL) to the Smallfoot Intermediate Language interpreted by Infer, thereby enabling Infer's capabilities on all CIL languages (including C\# and F\#). For the purposes of this paper, InferSharp refers to the static analysis of Infer applied to CIL languages. Notably, to our knowledge it is the only interprocedural static analysis for CIL languages which is free-to-use and MIT-licensed. Considering Infer's industry track record, this creates unique opportunities in both research and industry to build bug-detection-and-fix product capabilities for a relatively underserved developer segment.

\subsection{Collecting Data with Infer}

In this section we describe the data extraction process that culminated in the creation of the \inferredbugs~dataset. Specifically, we provide details on how we executed Infer over the change histories of software projects in order to detect introduced and fixed bugs.

Given as input the current commit \textit{curr} and the previous commit \textit{prev}, we begin by computing a \texttt{git diff} to identify the files involved in the change performed by the developer in the commit \textit{curr}. Next, we analyze the status of the files at commit $\textit{prev}$. Specifically, we checkout the snapshot of the system at commit $\textit{prev}$, and we build the system using the project-specific build tool. During the build process, the $\texttt{infer capture}$ command intercepts calls to the compiler to read source files and translates them into an intermediate representation which will allow Infer to analyze these files. Next, we invoke the $\texttt{infer analyze}$ command specifying the files to be analyzed (\ie the files $\textit{diff}$ involved in the commit). This analysis produces a report $\textit{reportPrev}$ detailing the bugs identified within the specified files.


Subsequently, we move to the current commit \textit{curr} and perform the same steps described for the commit \textit{prev}, that is: checking out the commit, building system while capturing the source files, and analyzing the \textit{diff} files in order to detect bugs. 

Finally, with the \texttt{infer reportdiff} command, we compute the differences between the two infer reports \textit{reportPrev} and \textit{reportCurr}. The output \textit{bugs} contain three categories of issues:
\begin{itemize}
    \item \textit{introduced}: issues found in \textit{curr} but not in \textit{prev};
    \item \textit{fixed}: issues found in \textit{prev} but not in \textit{curr};
    \item \textit{preexisting}: issues found in both \textit{prev} and \textit{curr}.
\end{itemize}

  
  
  
  
  
  
  
  
  
  


We perform these steps for each pair of commits (\textit{prev, curr}) over the change histories of the analyzed software projects. We optimize this process by obviating the need to build the same commit twice (\ie once as \textit{curr} and next as \textit{prev}) by instead reusing the build and capture stages in the next iteration.

\subsection{Dataset Statistics}

After running the extraction pipeline on 2937 repositories, we identified a total of 8280 bug patches. Of these bugs, 259 of these are null dereference patches which pass the filtering process, and 462 of these are resource leaks which pass the filtering process. We note that the filtered dataset contains commits which might have been detected by traditional methods involving extracting commits with certain keywords related to the desire bug type. Of the 259 null patches, 59 contain ``null'' or ``npe'' in the corresponding commit message, and of the 462 resource leak patches, 15 contain the ``leak'' keyword. We see from this that we are able to extract many additional fixes that would not have appeared using naive commit message keyword matching.


\begin{table}[htb]
\small
\caption{Summary of the \inferredbugs{} dataset in terms of the number of files and size of the bug-fixing two-way diff.}
\centering
\begin{tabular}{llllllllllll} \toprule
& \multicolumn{2}{c}{\textbf{NPD}}& \multicolumn{2}{c}{\textbf{RL}}& \multicolumn{2}{c}{\textbf{TSV}}\\
\cmidrule{2-7}  
& \textit{Java} & \textit{C\#} & \textit{Java} & \textit{C\#} & \textit{Java} & \textit{C\#} \\
\midrule
Num. bug patches & 2686 & 1116 & 2382 & 1789 & 3582 & 40 & \\
Mean lines per patch & 12.2 & 8.8 & 10.9 & 7.2 & 14.1 & 17.1 \\
Mean char per patch & 457.1 & 310.2 & 404.1 & 275.8 & 482.7 & 455.3 \\
\bottomrule
\end{tabular}
\label{tab:dataset}
\end{table}
As shown in \autoref{tab:dataset} the \inferredbugs{} is composed of multi-line bugs, which represents a challenging case for program repair tools.

\section{Baselines}
\label{sec:baselines}

In the following, we explore several program repair baselines which are constructed around powerful LLMs (\texttt{code-cushman-001} and \texttt{text-davinci-003}) for tasks of completing code, filling code in the middle, or generating a fix following a natural language instruction. In the following, we evaluate the performance based on the accuracy of exact string match of a generated patch to the ground truth fix.

\subsection{Demonstration Prompting}
\begin{figure}
\includegraphics[width=0.95\columnwidth]{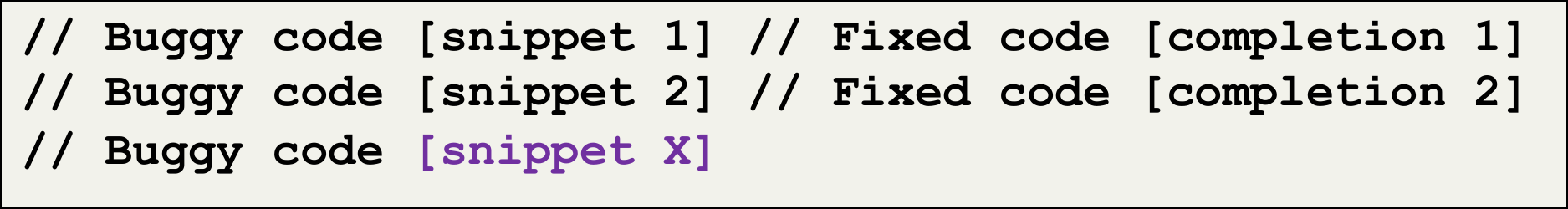}
    \caption{Demonstration prompt design for program repair experiments with LLMs.}
    \label{fig:demonstration}
\end{figure}

Demonstration learning is a prompt augmentation technique in which a few answered prompts are prepended to the context with the purpose of demonstrating how a language model should approach a downstream task. For program repair, we introduce a prefix constructed of two answered prompts as, followed by the actual buggy code snippet $\texttt{[X]}$, as shown in \autoref{fig:demonstration}. 
Our few-shot demonstration learning experiments are based on the strong 12 billion parameter Codex language model of code.  

\subsection{Conditional Language Modeling}
Our next baseline is the zero-shot conditional language generation (code completion), which aims to utilize the next token prediction to repair programs. Specifically, given a bug-free prefix, we run Codex model inference to complete the buggy code snippet, aiming to rewrite a program without bugs. In our experiments, we apply nucleus sampling decoding algorithm with $top\_p = 1$ and a temperature $T = 0.7$ generating top 10 samples up to the length of 1024 tokens with a total length for prefix and completion of 2048. Our conditional language modeling experiments are also based on the \texttt{code-cushman-001}.

\subsection{Instruction Prompting}
Instruction learning is a prompt augmentation technique that introduces a natural language description of the task. To approach program repair, we prepare prompts following a template:
\begin{figure}
\includegraphics[width=0.95\columnwidth]{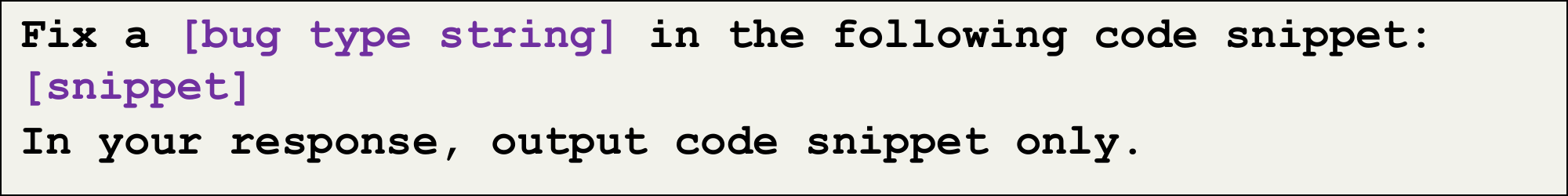}
    \caption{Instruction prompt design for program repair experiments with LLMs.}
    \label{fig:instruction}
\end{figure}
We utilize OpenaAI GPT-3 Davinci model, a 175 billion parameter language model and a close sibling of ChatGPT, to complete the prompts. Typically, Davinci outputs a natural language summary of the proposed fix followed by a code snippet. 
For the sake of evaluation, we instruct \texttt{text-davinci-003} to only output code snippet in its response which otherwise normally accompanied by the natural language descriptions.

\section{\thistool{} Framework}

\thistool{} program repair framework is composed of three following key modules: (i) a static analysis tool that detects, localizes, and classifies bugs, (ii) retrieval module – a large index of historic bugs and fixes, equipped with a facility to efficiently search and retrieve ``hints'' -- semantically-similar source code segments --  given a query, and (iii) generator module – a large language model finetuned on a dataset of prompts enriched with the information provided by the static analyzer and the retriever to generate fixes.

\subsection{Bug Detection \& Classification Module}
Our bug detection, localization, and classification module is powered by the Infer, which performs program analysis via Separation Logic. Although Infer's Pulse framework has recently been released, for the purposes of this paper we examine bugs generated by Infer's biabduction framework. Compiler frontends for Infer, such as InferSharp, translate source code into the control-flow-graph intermediate representation understood by Infer, known as the Smallfoot Intermediate Language. Infer performs automated program analysis over this graph and produces compositional method summaries in order to determine whether there are defects present in the source code.  

\subsection{Retrieval Module}

Our retrieval module closely follows the ReACC formulation~\cite{lu-etal-2022-reacc}. The retriever searches for semantically equivalent vulnerable code given a buggy code snippet and retrieves corresponding fix candidates based on cosine similarity between the embedding of query vector $q$ and a buggy code snippet $c$. 

Dense retrieval maps each code snippet to a $d$-dimension dense vector. The relevance of a code snippet to a given query can then be determined as a dot product of the query vector and each document vector. We closely follow the Dense Passage Retriever (DPR) model~\cite{karpukhin-etal-2020-dense}. At the training stage, we adopt in-batch negatives to calculate the contrastive loss by InfoNCE~\cite{Oord2018RepresentationLW}.

Our dense retriever utilizes a bidirectional transformer encoder $\mathcal{E}$ to obtain encoded dense vector representations of the query ($\mathcal{E}(q)$), and for each buggy code snippet $c$ indexed in the retrieval database ($\mathcal{E}(c)$). The retrieval database is a key-value store with encoded buggy code snippets $\mathcal{E}(c)$ serving as keys, and string representations of the corresponding fixes $f$ serving as values. 

We take the representation of the $\texttt{[CLS]}$ token as a summary of the encoded sequences of tokens, and compute similarity between the query and each code snippet in the database as a dot-product: $sim(q, c) = \mathcal{E}(c)^T \cdot \mathcal{E}(q)$. 

The bidirectional transformer encoder $\mathcal{E}$ is pretrained with the contrastive learning objective. Contrastive learning~\cite{Wu2018UnsupervisedFL,} is a self-supervised learning technique, in which the machine learning model is aiming to learn from the commonality of the training samples but also the attributes that make samples different. Given a contrastive pretraining dataset $D = \{q_i, p_i^+, p_{(i,1)}^-,...,p_{(i,h)}^-\}$, $i = 0...N$, where each sample consists of a query -- an encoding of a buggy code snippet; a positive sample representing a semantically similar code snippet of the same bug type; and a set of negative samples which are irrelevant code snippets of different bug types. The contrastive loss is then given by the following formula (negative log likelihood of the positive sample):  
\begin{equation}
L (q_i, p_i^+, p_{(i,1)}^-,...,p_{(i,n)}^-) = \\
- log \frac{e^{sim( q_i,p_i^+)}}{e^{sim( q_i, p_i^+)} + \sum_{i=1}^n e^{sim(q_i, p_{(i,j)}^-)}},
\end{equation}
where $sim$ is the cosine similarity between the embedding vectors.

\subsection{Generator Module}
Our generator model is based on Codex Cushman (\texttt{code-cushman-001}), a 12B parameter decoder-only transformer language model~\cite{codex} developed by OpenAI, which is a descendant of GPT-3, trained on source code.

We finetune Codex on a supervised corpus extracted from the \inferredbugs{} dataset, with the goal of teaching the model to generate a fix for the given buggy code. Specifically, the input to the model is the buggy code augmented with additional information such as bug localization and categorization, hierarchical extended context, and retrieved similar fixes. We discuss the prompt augmentation process in detail in \autoref{sec:prompt_augmentation}.

We perform full model finetuning (updating all weights of the model), on sixty four 32 GB V100 GPU for 5 epochs, retaining best model checkpoint by the exact match accuracy metric. We utilize Babel platform -- a model repository and an AzureML designer component family bringing together state-of-the-art transformer models on Azure ML compute for rapid experimentation. We use Adam stochastic optimization procedure with the learning rate of 0.01, warmup period of 1000 optimization steps, and global batch size of 256.

\section{Prompt Augmentation}
\label{sec:prompt_augmentation}

Prompt augmentation has been shown to be a powerful technique for extracting high-quality outputs from large language models, and, in particular, for domain and task adaptation. In the following we describe our dedicated prompt augmentation approach for program repair task. The proposed approach is two-fold: (i) we extract and prioritize syntax hierarchies which are most relevant to the buggy snippet region, including focal context, and (ii) retrieve hints -- structurally similar bug fixes from commit histories on GitHub. By doing so we are constructing a loosely structured template which includes the following: 
\begin{enumerate}
\item{Retrieved hints}
\item{Bug type annotation}
\item{Syntactic hierarchies and peer methods}
\item{Focal methods}
\item{Buggy method with location markers}
\end{enumerate}

\begin{figure*}
\includegraphics[width=0.95\textwidth]{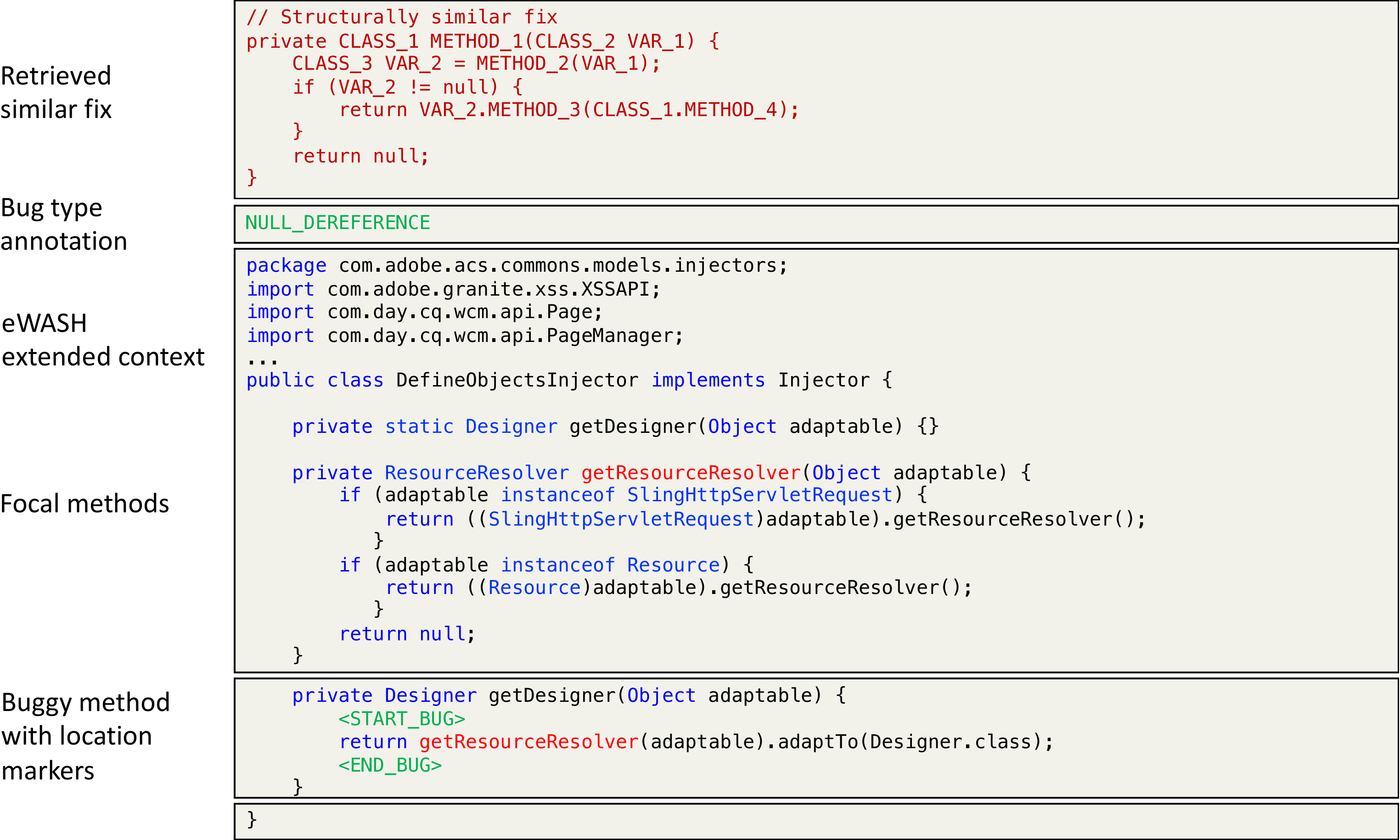}
    \caption{Prompt augmentation for a method in Java programming language affected by a null pointer dereference bug.}
    \label{fig:context}
\end{figure*}

\autoref{fig:context} shows an example of augmented prompt input for a null pointer dereference bug in Java, which includes the buggy code region surrounded by location markers, containing the method with surrounding most relevant syntax hierarchies, but type annotation string, and ``hints'' -- structurally similar bug fixes retrieved from the historic database. 

In the following subsections, we will describe each prompt augmentation technique and quantify its impact on bug-fixing performance by adding features incrementally.

\subsection{Basic Prompt}
The most basic prompt we can construct for the model is to provide the buggy method as input while expecting the model to generate the fix by outputting the fixed version of the given method. Thus, we perform task-oriented finetuning of our Generator model (Codex) using the buggy and fixed versions of the methods from the \inferredbugs{} dataset. 

We compare this basic prompting and finetuning against powerful LLM baselines described in Sec. \ref{sec:baselines}. \autoref{tab:basic_prompt} illustrates the effect of finetuning as compared to zero-shot and few-shot variants. Demonstration learning appears to be the most successful few-shot learning strategy for adapting the Codex model (\texttt{code-cushman-001}) to downstream patch generation task, yielding a modest 19--25\% accuracy of fixing Java bugs. Instruction learning, which also includes natural language descriptions of the downstream task in the prompt, only becomes viable as the model size increases -- we repeated the instruction learning experiments with the 175 billion parameter Davinci model (\texttt{text-davinci-003}), a close sibling of ChatGPT. We observe a very competitive performance with the Davinci variant, with 40--53\% accuracy of fixing Java bugs via prompt augmentation alone. Task oriented finetuning, without any prompt crafting, outperforms all few-shot baselines by a good margin, showing 11--55\% relative improvement of accuracy across all bug types in Java. This improvement comes at a cost of computing resources necessary to finetune the Codex model, but provide an advantage of higher accuracy and cheaper inference as compared to few-shot Davinci. 

\begin{table}[htb]
\small
\caption{Evaluation results for \thistool{} with basic prompt compared against LLM baselines}
\centering
\begin{tabular}{llllllllllll} \toprule
\textbf{Approach} & \multicolumn{2}{c}{\textbf{NPD}}& \multicolumn{2}{c}{\textbf{RL}}& \multicolumn{2}{c}{\textbf{TSV}}\\
\cmidrule{2-7}  
& \textit{Java} & \textit{C\#} & \textit{Java} & \textit{C\#} & \textit{Java} & \textit{C\#} \\
\midrule
Demonstration (Codex) & 20.3 & 30.1 & 25.3 & 29.1 & 19.0 & 16.7 \\
Completion (Codex) & 6.7 & 6.1 & 7.8 & 5.7 & 3.9 & 0.0\\
Instruction (Davinci) & 40.5 & 22.2 & 53.8 & 19.7 & 41.3 & 33.3 \\
\thistool{} (basic prompt) & 49.7 & 58.1 & 60.0 & 51.9 & 64.4 &  70.0\\
\bottomrule
\end{tabular}
\label{tab:basic_prompt}
\end{table}


\subsection{Bug Type Annotations}

The simplest prompt augmentation step is to prepend a bug type annotation to a basic prompt consisting of a buggy method only. As shown in \autoref{tab:nobt_ablation}, this improves performance across all bug categories and languages (Java and C\#) yielding 2.7--5.6\% relative improvement in accuracy.
\begin{table}[htb]
\small
\caption{Evaluation results for \thistool{} demonstrating the impact of introducing the bug-type annotation in the prompt.
}
\centering
\begin{tabular}{llllllllllll} \toprule
& \multicolumn{2}{c}{\textbf{NPD}}& \multicolumn{2}{c}{\textbf{RL}}& \multicolumn{2}{c}{\textbf{TSV}}\\
\cmidrule{2-7}  
& \textit{Java} & \textit{C\#} & \textit{Java} & \textit{C\#} & \textit{Java} & \textit{C\#} \\
\midrule
\thistool{} (basic prompt) & 49.7 & 58.1 & 60.0 & 51.9 & 64.4 &  70.0 \\
\thistool{} (+ bug type) & 52.3 & 60.4 &  63.1 & 53.3 & 67.9 & 72.5 \\
\bottomrule
\end{tabular}
\label{tab:nobt_ablation}
\end{table}

\subsection{Bug Localization}

Bug location information is crucial for accurate program repair. Infer static analyzer can localize bugs by tracking the flow of data through the program and detecting any violations of predefined rules or programming patterns. Infer static analyzer outputs a line number on which an error could occur at runtime, which, however, does not mean that the fix would require to edit this line only. In our dataset~\autoref{tab:dataset}, the bugs are often spanning over multiple lines of code, having disjoint diff regions.

We utilize the bug location information output by Infer in two ways: (i) we parse the source code file affected by the bug to extract a method which contains the buggy line, and (ii) we surround the buggy region with special sentinel $\texttt{<START\_BUG>}$ and $\texttt{<END\_BUG>}$ symbols. During training, we refine the bug location by looking at the two-way diff markers with respect to the fix. During test time, we only use the information provided by the static analyzer as the fix is unknown. 
\begin{table}[htb]
\small
\caption{Evaluation results for \thistool{} showing the effect of adding bug location markers in the prompt.}
\centering
\begin{tabular}{llllllllllll} \toprule
& \multicolumn{2}{c}{\textbf{NPD}}& \multicolumn{2}{c}{\textbf{RL}}& \multicolumn{2}{c}{\textbf{TSV}}\\
\cmidrule{2-7}  
& \textit{Java} & \textit{C\#} & \textit{Java} & \textit{C\#} & \textit{Java} & \textit{C\#} \\
\midrule
\thistool{} (bug type) & 52.3 & 60.4 &  63.1 & 53.3 & 67.9 & 72.5 \\
\thistool{} (+ localization) & 53.5 & 61.4 & 64.4 & 53.9 & 69.6 & 75.0  \\
\bottomrule
\end{tabular}
\label{tab:localization_ablation}
\end{table}
\autoref{tab:localization_ablation} demonstrates the impact of adding bug location markers in the prompt in addition to the bug type annotations. As seen, this leads to a positive effect across all categories studies, up to 3.4\% relative improvement in accuracy. The effect is more pronounced for larger methods.

\subsection{eWASH extended context}

A source code file may have nested scopes and references to other external libraries or other files.  To accurately suggest patches a model must leverage knowledge across different parts of the file. The length of source code files will often exceed the fixed-length window of transformer models (2048 tokens in our case), which could potentially lead to a loss of information relevant for learning to repair programs. To overcome this limitation, we utilize eWASH~\cite{clement-etal-2021-long} to prioritize syntax hierarchies which are most relevant to the buggy snippet region. Extracting syntactic hierarchies from the entire source code files, as opposed to the tokens immediately preceding the bug location, we are able to retain most relevant code context, such as class-level fields and method arguments, and peer methods which are highly relevant to program repair. Starting with a concrete syntax tree of a source file, we organize and prioritize class-level and method-level syntactic elements such as global import statements and assigned values, class attributes, method signatures, class docstring, and global expressions in the input. 

Quite often, a method affected by a bug will only contain an invocation expression or a call to a method defined elsewhere in the file – what we refer to as buggy \textit{focal method}. For instance, in \autoref{fig:context} the buggy line of code has a return statement which is composed of a chain of method invocations with \texttt{getResourceResolver} and \texttt{adaptTo} focal methods. We conjecture that retaining the focal method implementation (signature, docstring, and body) in the prompt is crucial for program repair. We utilize stack trace provided as part of the Infer bug report to determine relevant focal method name, and include it in the prompt. \autoref{tab:ewash_ablation} shows the effect of adding the eWASH syntax hierarchies and focal context in the prompt. As seen, patch generation accuracy is further improved by over 7.2--7.8\% for Java and by 4.0--6.7\% for C\#. 
\begin{table}[htb]
\small
\caption{Evaluation results for \thistool{} showing the effect of adding eWASH extended context in the prompt.}
\centering
\begin{tabular}{llllllllllll} \toprule
& \multicolumn{2}{c}{\textbf{NPD}}& \multicolumn{2}{c}{\textbf{RL}}& \multicolumn{2}{c}{\textbf{TSV}}\\
\cmidrule{2-7}  
& \textit{Java} & \textit{C\#} & \textit{Java} & \textit{C\#} & \textit{Java} & \textit{C\#} \\
\midrule
\thistool{} (localization) & 53.5 & 61.4 & 64.4 & 53.9 & 69.6 & 75.0 \\
\thistool{} (+ eWASH) & 57.6 & 65.1 & 69.1 & 56.1 & 75.0 & 80.0 \\
\bottomrule
\end{tabular}
\label{tab:ewash_ablation}
\end{table}

\subsection{Enriching Context with Hints}

To further enrich prompts, we perform a nearest neighbor search in the retrieval database for semantically similar fixes -- so called hints. The resulting fixes are then prepended to the context with an instruction string $\texttt{// Structurally similar fix}$. 

By default, we extract and prepend 2 nearest neighbors for each query. We apply quality criteria to avoid obviously incorrect matches: (i) retrieved fixes must be of the same bug type as the query, and (ii) impose a minimum similarity threshold between retrieved fixes and a query of 60\%.

\begin{figure}[htb]
\includegraphics[width=0.46\textwidth]{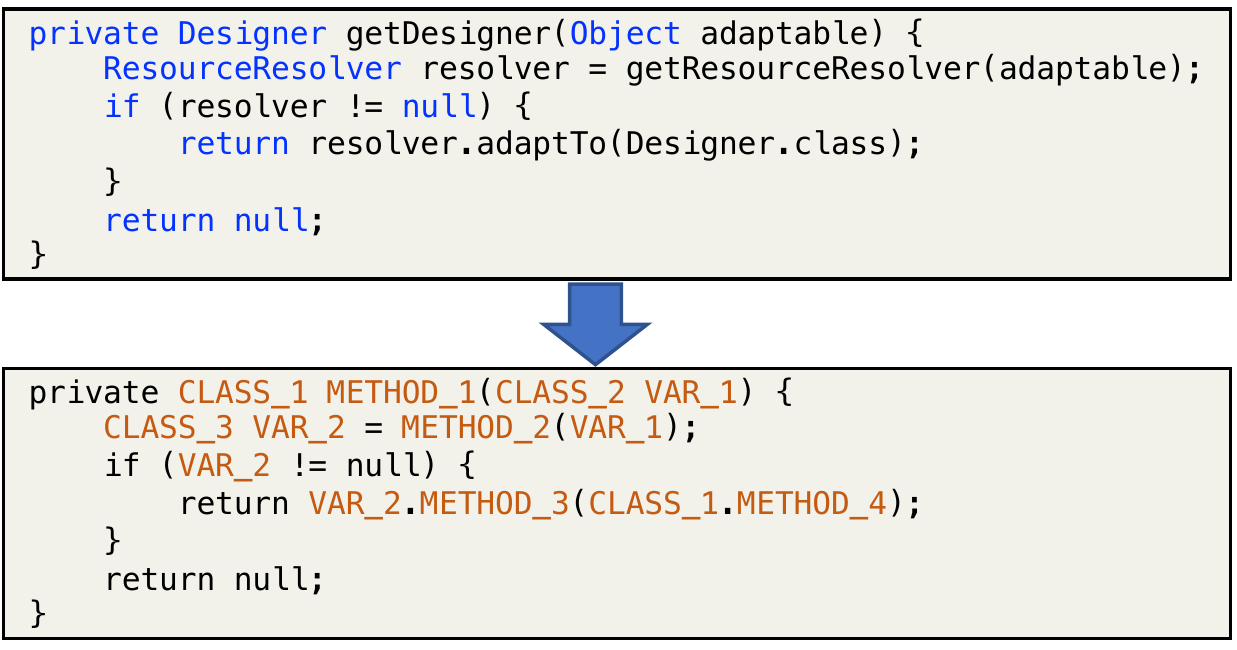}
    \caption{Code obfuscation example in Java.}
    \label{fig:obfuscate}
\end{figure}

To focus on extracting structurally similar fixes and reduce the dependency on identifier naming we obfuscate code snippets serving as keys in the database and search queries. Namely, we parse and analyze the code identifier types and mask the names of classes, methods, and identifiers with placeholder symbols: $\texttt{CLASS\_NN}$, $\texttt{METHOD\_NN}$, and $\texttt{VAR\_NN}$, where $\texttt{NN}$ is a unique number. An example obfuscated representation is shown in \autoref{fig:obfuscate}.

Table \ref{tab:hints_ablation} shows the improvements in bug-fixing capabilities for \thistool{} with prompt which incorporates retrieved hints. This prompt augmentation further improves \thistool{} performances by 1--2\% in absolute top-1 performances.

\begin{table}[htb]
\small
\caption{Evaluation results for \thistool{} showing the effect of adding bug-fix hints.}
\centering
\begin{tabular}{llllllllllll} \toprule
& \multicolumn{2}{c}{\textbf{NPD}}& \multicolumn{2}{c}{\textbf{RL}}& \multicolumn{2}{c}{\textbf{TSV}}\\
\cmidrule{2-7}  
& \textit{Java} & \textit{C\#} & \textit{Java} & \textit{C\#} & \textit{Java} & \textit{C\#} \\
\midrule
\thistool{} (eWASH) & 57.6 & 65.1 & 69.1 & 56.1 & 75.0 & 80.0 \\
\thistool{} (+ retrieved hints) & 59.5 & 66.7 & 71.2 & 57.0 & 77.4 & 82.5 \\
\bottomrule
\end{tabular}
\label{tab:hints_ablation}
\end{table}

\subsection{Inference}
The Inference step for \thistool{} involves utilizing nucleus sampling decoding with a top\_p parameter of 1.0 and a temperature of 0.7. During this step, the tool decodes the top-10 best predictions generated by the large language model, and ranks them according to their sequence log probabilities. This ranking helps to ensure that the most likely and relevant fixes are presented to the user. The use of nucleus sampling decoding, with its specific top\_p and temperature parameters, helps to balance the trade-off between diversity and quality in the generated predictions, making it possible to obtain highly accurate and diverse patch candidates.

\section{Results}
\autoref{tab:all_results} shows the results achieved by \thistool{} on the \inferredbugs{} dataset compared against the LLM baselines discussed in \autoref{sec:baselines}. \thistool{} is able to fix between 57\% and 82\% of the three categories of bugs for Java and C\#, just with the top-1 prediction. The performance gap between our approach and the best performing baseline (Finetuned Codex) is between 8.6\% and 13\% in absolute terms. 

It is important to keep in mind that the results shown in \autoref{tab:all_results} present a conservative estimate of \thistool{} 's potential for generating fixes. The percentages displayed in the table are based on generated patches that exactly match the original developer's token-by-token fix. However, there may be other candidate patches that correctly fix the bug using a different token sequence.

The impressive top-1 results achieved by \thistool{} are critical for its efficient and effective integration into the software development cycle. With the ability to propose high-quality fixes for critical bugs, \thistool{} has the potential to greatly enhance the productivity and reliability of the software development process.

\begin{table}[htb]
\small
\caption{Evaluation results for \thistool{} on the \inferredbugs{} dataset compared against LLM baselines}
\centering
\begin{tabular}{llllllllllll} \toprule
\textbf{Approach} & \multicolumn{2}{c}{\textbf{NPD}}& \multicolumn{2}{c}{\textbf{RL}}& \multicolumn{2}{c}{\textbf{TSV}}\\
\cmidrule{2-7}  
& \textit{Java} & \textit{C\#} & \textit{Java} & \textit{C\#} & \textit{Java} & \textit{C\#} \\
\midrule
Demonstration (Codex) & 20.3 & 30.1 & 25.3 & 29.1 & 19.0 & 16.7 \\
Completion (Codex) & 6.7 & 6.1 & 7.8 & 5.7 & 3.9 & 0.0\\
Instruction (Davinci) & 40.5 & 22.2 & 53.8 & 19.7 & 41.3 & 33.3 \\
Finetuning (Codex) & 49.7 & 58.1 & 60.0 & 51.9 & 64.4 &  70.0\\
\thistool{} & \textbf{59.5} & \textbf{66.7} & \textbf{71.2} & \textbf{57.0} & \textbf{77.4} & \textbf{82.5} \\
\bottomrule
\end{tabular}
\label{tab:all_results}
\end{table}

\section{Deployment}
The deployment of \thistool{} at Microsoft as part of the Azure DevOps and GitHub continuous integration pipeline (CI) has significantly improved the software development workflow for our internal projects at Developer Division. Such tight integration has enabled our software development teams to automate the bug detection and fixing process, reducing the time and effort required to manually identify and fix bugs, and ensuring that bugs are addressed quickly and accurately. \autoref{fig:pipeline} provides an overview of our CI pipeline with the integrated \thistool{} stages. When a pull request proposing code changes is created, the CI pipeline automatically triggers three steps: (i) build, (ii) testing, and (iii) Infer (or InferSharp) static analysis. If bugs are detected, the \thistool{} patch generation module is invoked to propose a fix. \thistool{} leverages the detailed information about the bug provided by Infer, such as context, location, and classification of the bug type.

The \thistool{} module proposes a (configurable) set of candidate patches. Each candidate patch is packaged as a separate Pull Request, which is individually validated. The validation process is seamless and reuses the three CI pipeline steps mentioned above. Specifically, the PR containing the candidate patch is validated through: (i) build -- checking that the candidate patch is syntactically and semantically correct w.r.t. the source project; (ii) testing -- ensuring that the candidate patch does not introduce regressions (failing tests); (iii) Infer static analysis -- validating that the candidate patch actually fixes the previously detected bug. The validated fix is then provided to the developer within the feature branch of the developer's Pull Request. The complexity of these stages are abstracted away from the developer, who will simply receive a PR comment within the system they are using (\eg GitHub or Azure DevOps). We implemented a GitHub action which receives a validated patch from \thistool{} and surfaces it to the developer in form of a GitHub comment in the PR. The comment provides detailed information about the bug (\ie extracted by Infer), and the resolution (\ie served by \thistool{}). The developer has the option to accept or decline the recommended fix.

The deployment of \thistool{} in the CI pipeline for our internal projects has provided significant benefits. Our software development teams can now focus on more important tasks, confident in the knowledge that bugs are being detected and fixed in a timely and efficient manner. We are currently in the process of expanding the number of projects that integrate \thistool into their CI pipeline, and the benefits of this integration have been demonstrated through the seamless validation process and abstracted complexity for the developer.

\section{Related Work}

Our approach is related to a broad set of literature on patch generation and prompting and task-oriented finetuning. We refer a reader to~\cite{repair_overview} for a more comprehensive overview on the prior research in the area of program repair, and~\cite{prompting_review} for a systematic survey of prompting methods in NLP.

Patches in the Wild~\cite{tufano_patches_wild}, utilize supervised machine translation to learn bug-fixing patterns for various common code defects. They mine bug-fixes from the change histories of projects hosted on GitHub and define the learning task on a method level, disregarding the surrounding code context. SequenceR~\cite{Chen2021SequenceRSL} improves upon the Patches in the Wild by leveraging the extended context available through the source code file containing the buggy code, showing first attempt at prompt crafting. SequenceR learning objective is based around supervised machine translation with encoder-decoder recurrent neural network. Copy That!~\cite{Panthaplackel2021CopyTE} builds upon an observation that patches typically only affect isolated spans of tokens, leaving most tokens unchanged. By introducing a span copying decoder they improve results upon the previous state-of-the-art. While also utilizing neural machine translation, DeepDebug~\cite{drain} leverage extensive self-supervised pretraining to improve upon the prior art. BugLab~\cite{NEURIPS2021_ea96efc0} takes a step towards self-supervised bug detection and repair, co-training two neural models: a detector model that learns to detect and repair bugs in code, and a selector model that learns to create buggy code for the detector to use as training data. CODIT~\cite{chakraborty2020codit} uses a tree-based model to encode source code changes, learning bug-fixing activities. Recoder~\cite{zhu2021syntax} generates edits in a syntax-guided manner and with a provider/decider architecture and placeholder generation. Lutellier \etal~\cite{lutellier2020coconut} employed ensemble learning with CNNs and NMT to generate patches with CoCoNuT. DLFix~\cite{li2020dlfix} is a two-tier model with the first layer focusing on learning the context of bug fixes and the second layer trying to generate the bug-fixing patch. Recently CURE~\cite{jiang2021cure} has reported state-of-the-art results on Defects4J and QuixBugs datasets, improving over NMT-based APR techniques with the use of a pre-trained programming language model, code-aware search, and sub-word tokenization. 

These works are trained on generic, unclassified bugs mined from change histories of open source repositories, and do not utilize the bug type information during learning. Differently, our proposed approach take advantage of the close relationship with the Infer static analyzer tool and leverages the bug type information during the learning process to generate specific fixes tailored for that category of bugs. Additionally, none of these aforementioned papers attempted to capitalize on large language models, as well as the effectiveness of prompt augmentation methods in connections to LLMs, combined with task-oriented finetuning.  

Our work is also aligned with a category of research that examines pretraining strategies and prompt augmentation. \cite{incoder} permute ordering of the spans in the original prompt to train the model to infill. Specifically, by randomly replacing spans of code with a sentinel token and moving them to the end of the sequence they yield a unified approach for both program synthesis (via left-to-right generation) and editing (via infilling). \cite{petroni-etal-2019-language} introduce a seminal LAMA dataset providing manually curated cloze templates to probe knowledge in language models. \cite{cui-etal-2021-template} investigate a template-based method for exploiting the few-shot learning potential of generative pre-trained language models to sequence labeling. Specifically, they define templates such as “$\texttt{<candidate\_span>}$ is a $\texttt{<entity\_type>}$ entity”, where $\texttt{<entity\_type>}$ can be “person” and “location”, etc, and train a model using a filled template. \cite{chain_of_thought} introduce a concept of chain of thought prompting, in which a task is broken down into a series of intermediate reasoning steps which significantly improves the ability of large language models to perform complex reasoning

Our proposed approach, \thistool, performs prompt augmentation by incorporating similar fixes identified in a historical database of bugs, along with other information. The concept of leveraging similar fixes has also been explored in other approaches, such as SimFix~\cite{jiang2018shaping}, which extracts frequent abstract modifications from existing patches to form an abstract space for program repair. It then analyzes similar code snippets in the same program to extract concrete modifications, which forms a concrete space. The intersection of these two spaces is used to perform fine-grained code adaptation for patch generation. Differently from the AST-differencing approach proposed in SimFix, we rely on a dense retrieval model which allows for more flexibility in identifying similar code snippets with arbitrary length, not constrained by specific AST-subtrees. Furthermore, our approach enhances the prompt by providing additional information and cues to the LLM model to facilitate the repair process.

\section{Conclusion}

We introduced \thistool{}: an end-to-end program repair framework based on Codex and a state-of-the-art static analyzer designed to fix critical security and performance bugs in Java and C\#. \thistool{} is based on a retrieval-based prompt augmentation technique and task-oriented finetuning that leverages bug-type annotations and extended source code context. We have also curated a \inferredbugs{}, a novel, metadata-rich dataset of bugs extracted by executing the Infer and InferSharp static analyzers on the change histories of thousands of Java and C\# repositories. Our experiments demonstrated that \thistool{} outperforms strong LLM baselines, reaching a top-1 accuracy of 65.6\% for generating fixes in C\# and 76.8\% in Java on the \inferredbugs{} dataset. 

We deployed \thistool{} internally at Microsoft as a GitHub action and as an Azure DevOps plugin operating as part of the continuous integration pipeline. This tool has significantly improved the software development workflow for our internal projects at Developer Division.


\bibliographystyle{ACM-Reference-Format}
\bibliography{bibliography}

\end{document}